\documentclass[aps,amsmath,onecolumn, showpacs,superscriptaddress,pre,10.5pt,nofootinbib]{revtex4-1}
\usepackage{amssymb}
\usepackage{amsmath}
\usepackage{graphicx}
\usepackage{array}
\usepackage{physics}
\usepackage{mathtools}
\usepackage{amssymb}
\usepackage{verbatim}
\usepackage{color}
\usepackage{bm}
\usepackage{bbm}
\usepackage{amsthm}
\usepackage[normalem]{ulem}
\usepackage[titletoc,title]{appendix}

\usepackage{fixmath}
\newcommand{\vect}[1]{\mathbold {#1}} 

\usepackage{color}
\usepackage[dvipsnames]{xcolor}
\definecolor{Blue}{rgb}{0.00, 0.00, 1.00}
\definecolor{Red}{rgb}{1.00, 0.00, 0.00}
\definecolor{Green}{rgb}{0.00, 0.50, 0.00}

\usepackage[colorlinks=true, urlcolor=blue, anchorcolor=blue, citecolor=blue,filecolor=blue,linkcolor=blue,menucolor=blue]{hyperref}

\newcommand{\bea}{\begin{eqnarray}}
\newcommand{\eea}{\end{eqnarray}}
\newcommand{\be}{\begin{equation}}
\newcommand{\ee}{\end{equation}}

\newcommand{\bee}{\begin{equation*}}
\newcommand{\eee}{\end{equation*}}

\colorlet{Mycolor1}{green!10!orange!90!}

\def\XXint#1#2#3{{\setbox0=\hbox{$#1{#2#3}{\int}$}
     \vcenter{\hbox{$#2#3$}}\kern-.5\wd0}}

\begin{document}
\title{Optimal finite-differences discretization for the diffusion equation from the perspective of large-deviation theory}

\author{Naftali R. Smith}
\email{naftalismith@gmail.com}

\affiliation{Department of Environmental Physics,
  Blaustein Institutes for Desert Research, Ben-Gurion University of
  the Negev, Sede Boqer Campus, 8499000, Israel}

\begin{abstract}

When applying the finite-differences method to numerically solve the one-dimensional diffusion equation, one must choose discretization steps $\Delta x$, $\Delta t$ in space and time, respectively. 
By applying large-deviation theory on the discretized dynamics,  we analyze the numerical errors due to the discretization, and find that the (relative) errors are especially large in regions of space where the concentration of particles is very small.
We find that the choice $\Delta t = {\Delta x}^2 / (6D)$, where $D$ is the diffusion coefficient, gives optimal accuracy compared to any other choice (including, in particular, the limit $\Delta t \to 0$), thus reproducing the known result that may be obtained using truncation error analysis. In addition, we give quantitative estimates for the dynamical lengthscale  that describes the size of the spatial region in which the numerical solution is accurate, and study its dependence on the discretization parameters.
We then turn to study the advection-diffusion equation, and obtain explicit expressions for the optimal $\Delta t$ and other parameters of the finite-differences scheme, in terms of $\Delta x$, $D$ and the advection velocity.
We apply these results to study large deviations of the area swept by a diffusing particle in one dimension, trapped by an external potential $\sim |x|$.
We extend our analysis to higher dimensions by combining our results from the one dimensional case with the locally one-dimension method.

\end{abstract}

\maketitle

\section{Introduction}

The diffusion equation
\be
\partial_{t}u=D\nabla^{2}u
\ee
and the advection-diffusion equation
\be
\label{AdvectionDiffusion}
\partial_{t}u+\vect{F}\cdot\nabla u=D\nabla^{2}u
\ee
(in $d$ spatial dimensions) are of central importance in many fields, including hydrodynamics, statistical physics, biology, engineering and many more.
In these equations, $u\left(\vect{x},t\right)$ is the concentration of some substance, $D>0$ is the diffusion coefficient and in Eq.~\eqref{AdvectionDiffusion}, $\vect{F}$ is the velocity due to advection.
In simple cases these equations can be solved analytically, but in general they can be difficult to solve, and as a result many numerical methods have been developed \cite{GW74, GG78, Iserles95, Pert81, ALV96, SH07, LP11, HBMS11, FB14, NNS16, SM17, SMR22, LY23, FPT23}.
One of the simplest methods involves a discretization of time and space, followed by an approximation of the temporal and spatial derivatives by using the method of finite differences. 
One is then faced with the problem of choosing the discretization steps $\Delta x$, $\Delta t$ in space and time, respectively. It is well known that if this choice is not done carefully, the numerical solution may be unstable even if $\Delta x$ and $\Delta t$ are both very small (see below). 

For simplicity, let us begin by discussing the diffusion equation in $d=1$,
\be
\label{diffusion1D}
\partial_{t}u=D\partial_{x}^{2}u \, .
\ee
We choose a rectangular grid in the $xt$ plane, defined by the points
\bea
&&x_{i}=i\Delta x,\quad i\in\mathbb{Z},\\
&&t_{j}=j\Delta t,,\quad j=0,1,2,\dots,
\eea
and denote the concentration at position $x_i$ and time $t_j$ by $u_{i}^{j}$.
Approximating the derivatives by using the method of finite differences, one obtains
\be
\frac{u_{i}^{j+1}-u_{i}^{j}}{\Delta t}=D\frac{u_{i+1}^{j}-2u_{i}^{j}+u_{i-1}^{j}}{\Delta x^{2}} \, ,
\ee
which can be rewritten as
\be
\label{RW1D}
u_{i}^{j+1}=u_{i}^{j}+p\left(u_{i+1}^{j}-2u_{i}^{j}+u_{i-1}^{j}\right)
\ee
where $p=D\Delta t/\Delta x^{2}$.
In principle, one must also deal with the boundary conditions. However, for now they will not play an important role in our analysis so we will assume for simplicity that the system size is infinite.

What is the best way to choose the space-time discretization?
Generally speaking, as $\Delta x$ and $\Delta t$ tend to zero, the accuracy of the solution improves but the computational cost increases.
However, it is well known that, for the numerical solution to be stable, one must choose $p \le 1/2$, see e.g. \cite{Iserles95}.
Phrasing the question slightly differently, for a given spatial discretization $\Delta x$, how should one choose $p$ (or equivalently $\Delta t$) to obtain good efficiency and accuracy?
One may naively expect that, as $p$ decreases, the accuracy of the solution improves but the computational cost increases.
However, it turns out that (rather surprisingly and counterintuitively) the best accuracy is obtained for $p=1/6$ and, in particular, if $p$ is decreased below this value then both the efficiency and the accuracy of the numerical solution worsen. This result is well-known and is easy to obtain using error-truncation analysis,  see e.g. Ref.~\cite{GG78} and references therein. For completeness we reproduce this analysis  in Appendix \ref{app:truncation}.

In this paper, we revisit this problem from the point of view of large-deviation theory. We first note that the relative numerical errors are especially large in regions of space in which the concentration of particles $u$ is very small. We characterize the region of space in which the relative numerical errors are small through a dynamical lengthscale $\ell(t)$. We find that, for generic choices of $0< p \le 1/2$, this lengthscale behaves (at $t \gg \Delta t$) as 
\be
\label{ellBad}
\ell(t) \sim t^{3/4},
\ee
but for the specific choice $p=1/6$, 
\be
\label{ellGood}
\ell(t) \sim t^{5/6},
\ee
an improvement that becomes especially significant at long times.

The remainder of the paper is organized as follows.
In section \ref{sec:Diffusion1D} we analyze the diffusion equation in $d=1$ in detail by applying tools from large-deviation theory.
In section \ref{sec:AdvectionDiffusion1D} we perform an analogous analysis for the one-dimensional advection-diffusion equation, which we then apply in section \ref{sec:Reflected} to numerically study large deviations of the area swept by a Brownian particle trapped in an external potential $\sim |x|$.
In section \ref{sec:highDim} we extend our analysis to $d>1$.
Finally, we briefly summarize and discuss our results in section \ref{sec:conclusion}.

\section{Diffusion equation in $d=1$}
\label{sec:Diffusion1D}

Our starting point is the diffusion equation \eqref{diffusion1D}.
In what follows, it is very useful to interpret this as a Fokker-Planck equation for the time-dependent probability density function $u(x,t)$ of the position of a single, diffusing particle (so that
$u\left(x,t\right)dx$ is the probability that, at time $t$, the position of the particle is between $x$ and $x+dx$) \cite{vanKampen, Gardiner}.
One can then interpret the discretized equation \eqref{RW1D} as describing the temporal evolution of the probability vector $u_{i}^{j}$ of the position of a particle undergoing a random walk in discrete space and time. Indeed, interpreting $u_{i}^{j}$ as the probability that, at time $t_j$ the position of the particle is $x_i$, the dynamics \eqref{RW1D} describe a particle which, at every timestep (of duration $\Delta t$), either jumps a distance of $\Delta x$ to the left or to the right, each with probability $p$, or does not move, with the complementary probability $1-2p$.
Incidentally, within this interpretation the requirement $p \le 1/2$ 
 corresponds to the probability of not moving being nonnegative.
An equivalent way to represent the dynamics of the random walk would be to write them as
\be
\label{RWdef}
X^{j+1}=X^{j}+\xi^{j} \Delta x \, ,
\ee
where $X^j$ is a random variable denoting the position of the particle at time $t_j$, and $\xi^{1},\xi^{2},\dots$ are i.i.d. random variables each taking the values $-1,0,1$ with probabilities $p,1-2p,p$ respectively.

How well does the discrete random walk approximate the (continuous) diffusion process?
For simplicity, let us assume that the particle starts from position $x=0$ at time $t=0$, i.e., the initial condition for Eqs.~\eqref{diffusion1D} and \eqref{RW1D} are given by Dirac-
$u\left(x,t=0\right)=\delta\left(x\right)$
and Kronecker-
$u_{i}^{0}=\delta_{i,0}$
delta functions, respectively.
Moreover, we will assume for now that the system is infinite in space, i.e., $-\infty < x < \infty$ in \eqref{diffusion1D} and $i \in \mathbb{Z}$ in \eqref{RW1D}.
In this case, the solution to \eqref{diffusion1D} is simply given by the Green's function of the diffusion equation,
\be
\label{Green}
u\left(x,t\right) = G\left(x,t\right) =\frac{1}{\sqrt{4\pi Dt}}e^{-x^{2}/\left(4Dt\right)} \, ,
\ee
to which we would now like to compare the solution to the discretized equation \eqref{RW1D}.

One can gain insight by using the representation \eqref{RWdef} of the random walk, which we analyze here with the initial condition $X^{0}=0$.
In this representation it is clear that
$X^{j}=\Delta x\sum_{j'=0}^{j-1}\xi^{j'}$,
and since the $\xi^{j'}$ are i.i.d., $X^{j}$ converges to a normal distribution at large $j$.
Moreover, since the mean and variance of $\xi^{j}$'s are given by
$\left\langle \xi^{j}\right\rangle =0$
(where the angular brackets denote ensemble averaging) and
$\text{Var}\left(\xi^{j}\right)=2p$
respectively, it immediately follows that
$\left\langle X^{j}\right\rangle =0$
and
$\text{Var}\left(X^{j}\right)=2pj\Delta x^{2}=2Dt_{j}$.
Thus, at $j \gg 1$, typical fluctuations of $X^j$ are well approximated by a Gaussian distribution
\be
\label{RWGaussApprox}
P\left(X^{j}\right)\simeq\frac{1}{\sqrt{4\pi Dt_{j}}}e^{-\left(X^{j}\right)^{2}/\left(4Dt_{j}\right)} \, ,
\ee
(due to the central limit theorem) in agreement with Eq.~\eqref{Green}.

We are now interested in analyzing the error in the approximate result \eqref{RWGaussApprox}.
This can be done by considering the higher cumulants of the distribution of $X^j$: For an (exactly) Gaussian distribution, all cumulants higher than the second cumulant vanish. 
At long times, the nonvanishing higher cumulants of $P(X^j)$ mostly affect the tails of the distribution. Indeed, at large $j$, the large-fluctuations regime of this distribution follows the scaling behavior
\cite{O1989, DZ, Hollander, hugo2009, Jack20}
\be
\label{LDP}
P\left(X^{j}\right)\sim e^{-jI\left(X^{j}/j\Delta x\right)}\,.
\ee
This scaling behavior is known as a large-deviation principle, and it describes an exponential decay in time with a ``rate function'' $I(a)$.
The rate function vanishes at its global minimum which is at $a=0$, and its asymptotic behavior at $|a| \ll 1$ is
$I\left(a\right)\simeq a^{2}/4p$,
thus providing a smooth matching with the typical-fluctuations Gaussian regime \eqref{RWGaussApprox}.
$I(a)$ is closely related to the higher cumulants of the distribution: It is given by the Legendre-Fenchel transform 
\be 
I\left(a\right)=\sup_{k}\left\{ k a-\lambda\left(k\right)\right\} 
\ee
of the scaled cumulant generating function (SCGF) 
$\lambda\left(k\right)=\lim_{j\to\infty}\frac{1}{j}\ln\left\langle e^{kX^{j}/\Delta x}\right\rangle $ \cite{hugo2009}.
Due to the mirror symmetry of $P(X^j)$, the leading-order correction to the quadratic approximation of $I(a)$ around its minimum is generically of order $O(a^4)$.


$\lambda(k)$ can be easily found explicitly:
\be
\lambda\left(k\right)=\ln\left\langle e^{k\xi^{1}}\right\rangle =\ln\left(1-2p+2p\cosh k\right).
\ee
Writing this as a power series in $k$, we find
\be
\label{lambdaSolDiffusion}
\lambda\left(k\right)=pk^{2}+\frac{p\left(1-6p\right)}{12}k^{4}+O\left(k^{6}\right) \, .
\ee
 The power series expansion of $\lambda(k)$, taken up to quadratic order, describes the mean and variance of the Gaussian distribution of $X^j$ at $j \gg 1$, in agreement with the corresponding quadratic asymptotic behavior of $I(a)$ around its minimum (at $a=0$). The terms in the series \eqref{lambdaSolDiffusion} that are of order $O(k^n)$ with $n>2$ represent corrections to this Gaussian behavior [in particular, the $O(k^4)$ term in \eqref{lambdaSolDiffusion} gives rise to the $O(a^4)$ term in the expansion of $I(a)$], and their importance decreases as $n$ grows.
One immediately notices that the $O(k^4)$ term in Eq.~\eqref{lambdaSolDiffusion} term vanishes at $p=1/6$. I.e., for the choice $p=1/6$, the fourth cumulant of $X^j$ exactly vanishes and the subleading corrections to the leading-order Gaussian behavior come from the sixth cumulant and higher.
Thus, from the point of view of approximating the diffusion process by the random walk, the choice $p=1/6$ is expected to provide the best accuracy.

\begin{figure*}[ht]
\includegraphics[width=0.48\linewidth,clip=]{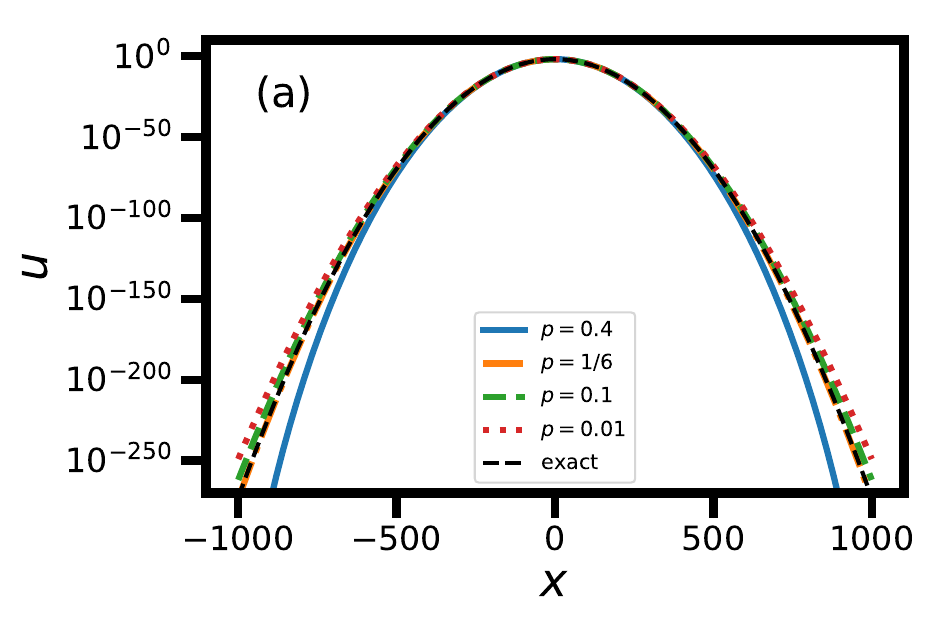}
\hspace{1mm}
\includegraphics[width=0.45\linewidth,clip=]{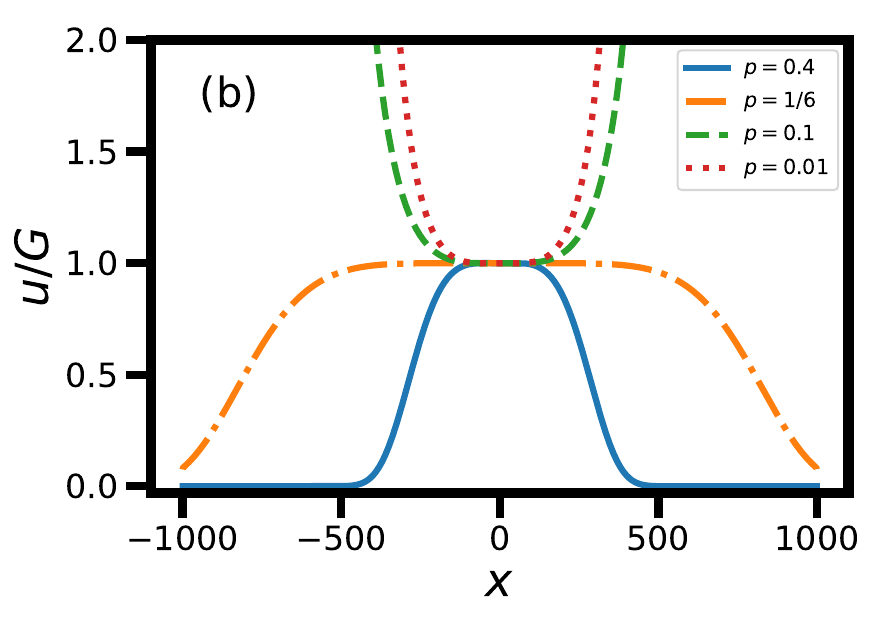}
\caption{(a) Numerical solutions $u(x,t)$ of the diffusion equation obtained using the finite-differences method, with $\Delta t= p \Delta x^2 / D$. Parameters are $D=\Delta x = 1$, $t=400$, and the different curves correspond to different values of $p$ as indicated in the figure label. The exact, analytic solution $G\left(x,t\right)$ \eqref{Green} is also plotted. (b) The ratio $u/G$ between the numerical solutions and the exact solution. It is clearly seen in the figure that for $p=1/6$, the agreement is significantly better than any other $p$.}
\label{fig:RWs}
\end{figure*}

We tested this prediction by solving Eq.~\eqref{RW1D} with different values of $p = 0.4, 1/6, 0.1, 0.01$, see Fig.~\ref{fig:RWs}. Indeed, $p=1/6$ provides a level of accuracy that is far better than all other choices.
The differences are especially pronounced in the tails of the distribution, which are overestimated (underestimated) for $p<1/6$ ($p>1/6$) by orders of magnitude.
 For example, for the choice $p=1/6$, $u\left(x=500,t=400\right)$ differs by just $4\%$ from its theoretical value $G\left(x=500,t=400\right)\sim{10}^{-70}$, whereas for the choices $p = 0.4, 0.1, 0.01$ the ratio $u/G|_{x=500,t=400}$ takes the values $3.6 \times {10}^{-4}, 6.5$ and $71$, respectively.

The large-deviation principle \eqref{LDP} enables us to estimate the range of $X^j$'s beyond which the numerical solution is expected to develop very large errors with respect to the exact solution, for different choices of the parameter $p$.
Indeed, for $p \ne 1/6$ the $O(a^4)$ term in the power-series expansion of $I(a)$ causes the numerical solution to lose accuracy at
$X^{j}\sim\Delta xj^{3/4}\sim\left(t_{j}D\right)^{3/4}/\left(\Delta x\right)^{1/2}$,
corresponding to Eq.~\eqref{ellBad} above.
In contrast, for $p=1/6$ the $O(a^4)$ term is absent, and instead the leading-order correction to the parabolic behavior comes from an $O(a^6)$ term. This causes the numerical solution to be accurate in a much wider range, losing accuracy only at
$X^{j}\sim\Delta xj^{5/6}\sim\left(t_{j}D\right)^{5/6}/\left(\Delta x\right)^{2/3}$, corresponding to Eq.~\eqref{ellGood} above.
The differences between these lengthscales is clearly seen in Fig.~\ref{fig:RWs}.

\section{Advection-diffusion equation in $d=1$}
\label{sec:AdvectionDiffusion1D}

Let us now consider the case in which a (constant) advection term is added, i.e., the equation of motion reads
\be
\label{diffusionDrift1D}
\partial_{t}u+F\partial_{x}u=D\partial_{x}^{2}u\,.
\ee
A standard approach for discretizing time and space would be to approximate the terms $\partial_{t}u$ and $\partial_{x}^{2}u$ as described above, and to approximate the advection term using the finite-differences method by
$\partial_{x}u\simeq\left(u_{i+1}^{j}-u_{i-1}^{j}\right)/\left(2\Delta x\right)$.
If this is done, Eq.~\eqref{RW1D} gives way to
\be
\label{RWDrift1D}
u_{i}^{j+1}=\left(1-p-q\right)u_{i}^{j}+qu_{i+1}^{j}+pu_{i-1}^{j}\,,
\ee
where
\be
\label{pqdef}
p=\left(\frac{D}{\Delta x^{2}}+\frac{F}{2\Delta x}\right)\Delta t,\quad q=\left(\frac{D}{\Delta x^{2}}-\frac{F}{2\Delta x}\right)\Delta t\,.
\ee
Once again one is faced with the problem of choosing the discretization parameters, which we shall phrase in the form of choosing $\Delta t$ for a given $\Delta x$.

We will follow the same general approach that we used above for the advection-free case.
We interpret Eq.~\eqref{RWDrift1D} as the evolution of a random walker in discrete time and space, that at every timestep $\Delta t$, jumps a distance of $\Delta x$ to the right (left) with probability $p$ ($q$), or stays at the same position with the complementary probability $1-p-q$.
Note that, clearly from the definition of $q$ in \eqref{pqdef}, one must choose $\Delta x\le2D/F$, otherwise $q$ is negative.
It is again convenient to use the description \eqref{RWdef} of the random walker's dynamics, but now the $\xi^j$'s take the values $-1,0,1$ with probabilities $q,1-p-q,p$ respectively.

The exact solution to the advection-diffusion equation \eqref{diffusionDrift1D} with initial condition $u(x,t=0) = \delta(x)$ is
\be
\label{GreenDrift}
u\left(x,t\right)=G_{F}\left(x,t\right)=\frac{1}{\sqrt{4\pi Dt}}e^{-\left(x-Ft\right)^{2}/\left(4Dt\right)}\,.
\ee
We now analyze the error of the numerical solution, which again is expected to be most pronounced in the tails of the distribution $P(X^j)$. 
Let us begin by obtaining the central limit theorem's prediction, which should give $P(X^j)$ in the typical-fluctuations regime.
The mean and variance of each of the $\xi^j$'s are
\be
\left\langle \xi^{j}\right\rangle =p-q,\quad\text{Var}\left(\xi^{j}\right)=p+q-\left(p-q\right)^{2}
\ee
respectively, so it immediately follows that the mean and variance of $X^j$ are given by
\be
\label{MeanXj}
\left\langle X^{j}\right\rangle =\left(p-q\right)j\Delta x=Ft_{j}
\ee
and
\be
\label{VarXjNaive}
\text{Var}\left(X^{j}\right)=\left[p+q-\left(p-q\right)^{2}\right]j\Delta x^{2}=\left(2D-F^{2}\Delta t\right)t_{j} 
\ee
respectively.
The mean \eqref{MeanXj} is in agreement with the exact result \eqref{GreenDrift}, but the variance \eqref{VarXjNaive} corresponds to an effective diffusion coefficient that is slightly modified compared to the exact one, $D_{\text{eff}}=D-F^{2}\Delta t/2$. Although the difference is not large (and vanishes as $\Delta t \to 0$), its effect on the tails of the distribution $P(X^j)$ can be nonnegligible.

\begin{figure*}[ht]
\includegraphics[width=0.48\linewidth,clip=]{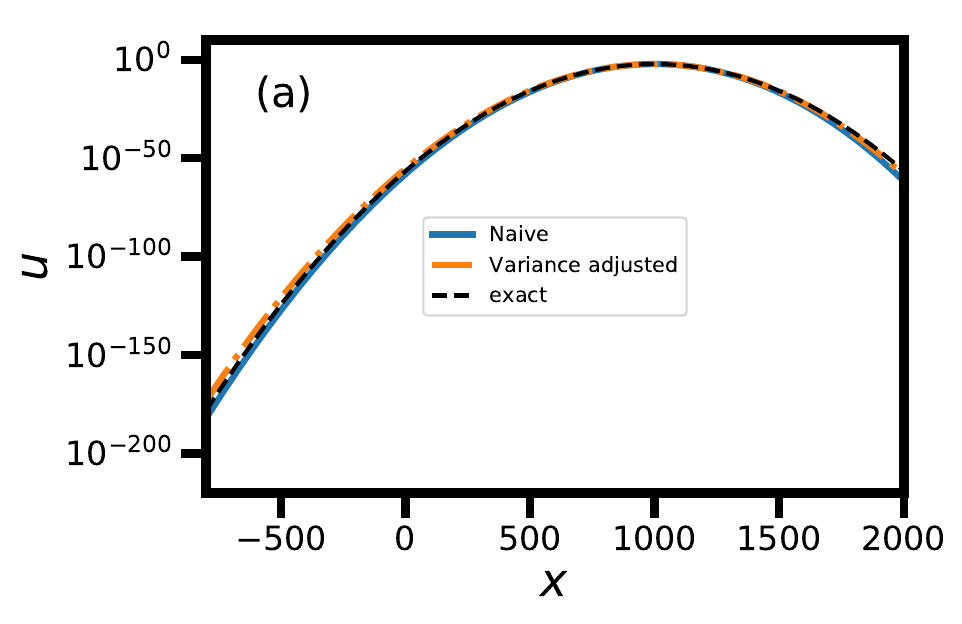}
\hspace{1mm}
\includegraphics[width=0.45\linewidth,clip=]{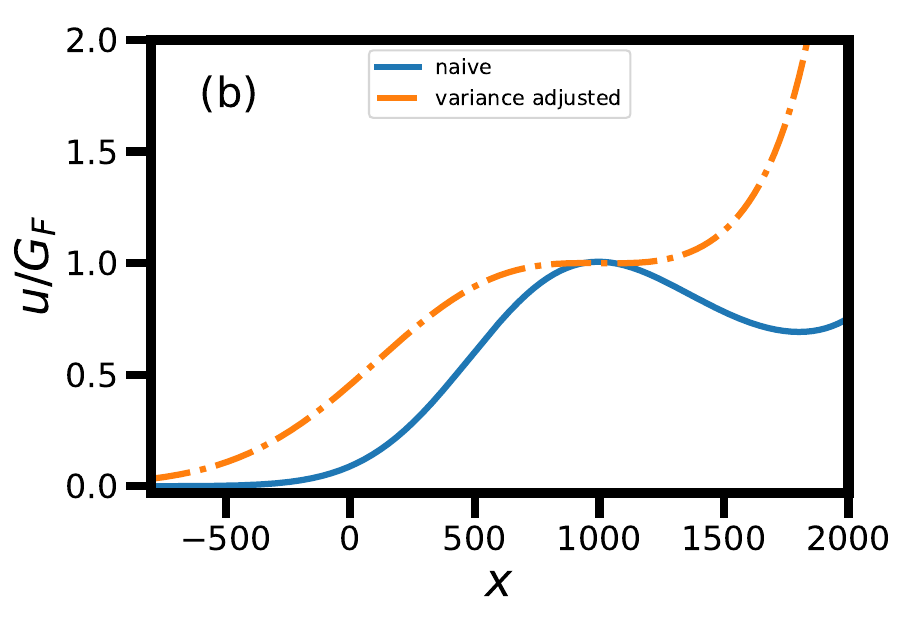}
\caption{(a) Numerical solutions $u(x,t)$ of the advection-diffusion equation obtained using the finite-differences method \eqref{RWDrift1D}, using the naive $p$ and $q$ from Eq.~\eqref{pqdef} (solid line), and using the improved values \eqref{pBetter} and \eqref{qBetter} (dot-dashed lines) that are adjusted so that the variance $X^j$ equals its theoretical value, together with the exact solution $G_F$ from Eq.~\eqref{GreenDrift}. Parameters are $D = \Delta x = 1$, $F=1/2$, $t=2000$ and $\Delta t = 0.1$.
(b) The ratio $u/G_F$ between the numerical solutions and the exact solution. As seen in the figure, the accuracy is somewhat improved when using the variance-adjusted $p$ and $q$.}
\label{fig:RWDsNaive}
\end{figure*}

We therefore propose, as a first improvement over this naive method, to modify the choices of $p$ and $q$ so that the mean and variance of $X^j$ attain their desired (exact) values. From the first parts of Eqs. \eqref{MeanXj} and \eqref{VarXjNaive}, one finds that the requirements are
\bea
p-q&=&\frac{F\Delta t}{\Delta x} \, ,\\
p+q-\left(p-q\right)^{2}&=&\frac{2D\Delta t}{\Delta x^{2}} \, .
\eea
The solution to these equations is
\bea
\label{pBetter}
p&=&\frac{1}{2}\left(\frac{2D\Delta t+F^{2}\Delta t^{2}}{\Delta x^{2}}+\frac{F\Delta t}{\Delta x}\right),\\
\label{qBetter}
q&=&\frac{1}{2}\left(\frac{2D\Delta t+F^{2}\Delta t^{2}}{\Delta x^{2}}-\frac{F\Delta t}{\Delta x}\right) ,
\eea
which replace the naive choices given by Eq.~\eqref{pqdef}.
In Fig.~\ref{fig:RWDsNaive}, we show that, for a particular set of parameters $(D,F,\Delta x)$, this choice of $p$ and $q$ indeed improves the accuracy of the solution, although the effect is not very large. We found similar improvements for other choices of parameters $(D,F,\Delta x)$ (not shown).

Having chosen $p$ and $q$ such that the mean and variance of $X^j$ are correct, one can now analyze the error of the numerical solution as encoded in the higher-order cumulants of $P(X^j)$.
As in the advection-free case, these are again related to the tails of the distribution as described by Eq.~\eqref{LDP}, but with a rate function $I(a)$ whose features are slightly modified, due to the advection, which breaks the mirror symmetry of the distribution $P(X^j)$.  As a result, $I(a)$ will also not be mirror symmetric around $a=a^*$.
So first of all, the minimum of $I(a)$ is at the nonzero value $a = a^* = F$, with an asymptotic behavior that is again parabolic
$I\left(a\right) \sim \left(a-a^{*}\right)^{2}$
 around $a \simeq a^*$.
Moreover, there will in general be a cubic term in this asymptotic expansion around $a=a^*$, reflecting the nonvanishing third cumulant of the distribution $P(X^j)$ [To remind the reader, the third cumulant for the exact (Gaussian) solution vanishes].

Nevertheless, as we now show, one can choose $\Delta t$ such that this third cumulant exactly vanishes, thus significantly reducing the error of the numerical solution.
The SCGF is again easy to calculate:
\be
\lambda\left(k\right)=\ln\left\langle e^{k\xi^{1}}\right\rangle =\ln\left(1-p-q+pe^{k}+qe^{-k}\right),
\ee
and its power-series expansion around $k=0$ is
\be
\lambda\left(k\right)=\left(p-q\right)k+\frac{1}{2}\left(p+q-p^{2}+2pq-q^{2}\right)k^{2}+\frac{1}{6}\left(p-q\right)\left(-4pq+2p^{2}-3p+2q^{2}-3q+1\right)k^{3}+O\left(k^{4}\right) \, .
\ee
From here, we read off the third cumulant
\be
\left\langle \left(\xi^{j}\right)^{3}\right\rangle _{c}=\left(p-q\right)\left(-4pq+2p^{2}-3p+2q^{2}-3q+1\right) \, .
\ee
Now, recalling the choices \eqref{pBetter} and \eqref{qBetter} of $p$ and $q$ (repsectively) in terms of $\Delta t$, we can determine the optimal choice of $\Delta t$ by requiring $\left\langle \left(\xi^{j}\right)^{3}\right\rangle _{c}$ to vanish, which (using that $F \ne 0$ so clearly $p \ne q$) leads to the equation
\be
\label{pqequation}
-4pq+2p^{2}-3p+2q^{2}-3q+1= 1-\frac{\left(6D+F^{2}\Delta t\right)\Delta t}{\Delta x^{2}} = 0
\ee
whose solution is%
\footnote{There is an additional solution to Eq.~\eqref{pqequation}, but is negative $\Delta t < 0$ so we discard it.}
\be
\label{DeltatSolRWD}
\Delta t = \Delta t^*=\frac{\sqrt{9D^{2}+\Delta x^{2}F^{2}}-3D}{F^{2}} \, .
\ee
This particular choice is expected to give the best accuracy, because for this choice the third cumulant of $P(X^j)$ vanishes so the error comes from higher cumulants.

 As in the case of the diffusion equation (with no advection), our choices for the parameters $p$, $q$ and $\Delta t$ coincide with the ones that may be obtained by truncation error analysis, see e.g. Ref.~\cite{GW74}. Besides reproducing the optimal values of these parameters, our analysis provides additional insight by interpreting the finite-differences method as the dynamics of a random walk in discrete space and time. Moreover, we point out that the numerical errors can be enormous (many orders of magnitude) in the large-deviation regimes, and find a dynamical lengthscale that describes the size of the region in space in which the relative numerical error is small.

\begin{figure*}[ht]
\includegraphics[width=0.48\linewidth,clip=]{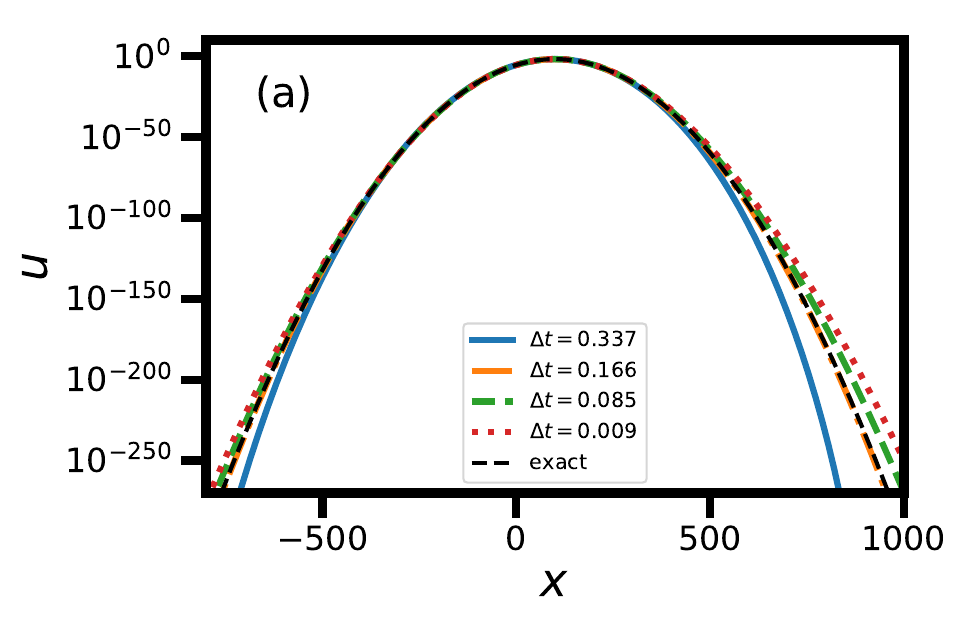}
\hspace{1mm}
\includegraphics[width=0.45\linewidth,clip=]{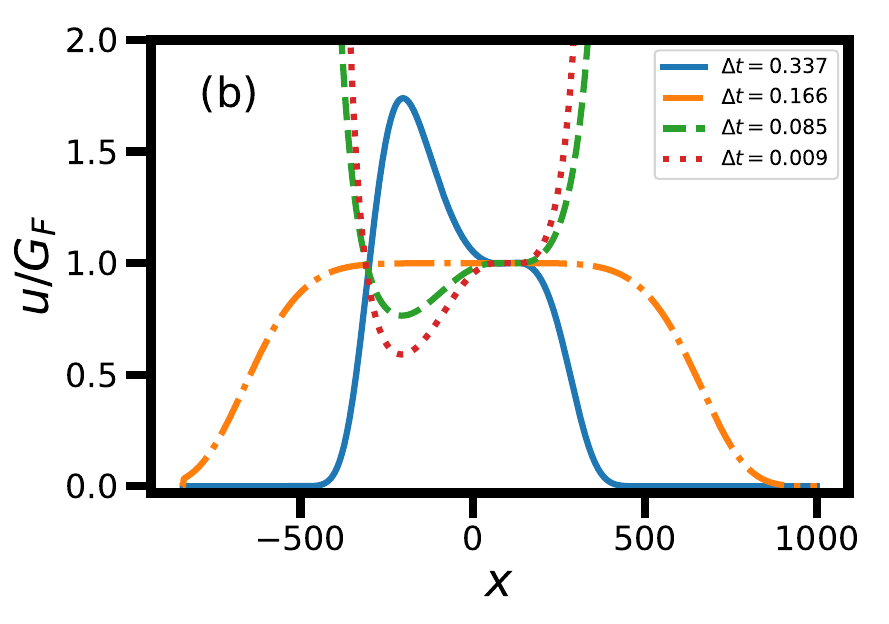}
\caption{(a) Numerical solutions $u(x,t)$ of the advection-diffusion equation obtained using the finite-differences method. Parameters are $D=\Delta x = 1$, $F=1/3$, $t=300$, for which Eq.~\eqref{DeltatSolRWD} gives $\Delta t^*  = 0.16615\dots$. The different curves correspond to different choices of $\Delta t$, as indicated in the figure label, from which the parameters $p$ and $q$ are determined via Eqs.~\eqref{pBetter} and \eqref{qBetter} respectively.
The exact, analytic solution $G_F\left(x,t\right)$ from Eq.~\eqref{GreenDrift} is also plotted. (b) The ratio $u/G_F$ between the numerical solutions and the exact solution. It is clearly seen in the figure that for $\Delta t = \Delta t^*$, the agreement is significantly better than any other $\Delta t$.}
\label{fig:RWDs}
\end{figure*}

We tested the quality of our choice of parameters
by solving Eq.~\eqref{RWDrift1D} with different values of $\Delta t$, see Fig.~\ref{fig:RWDs}. Indeed, $\Delta t = \Delta t^*$ provides a level of accuracy that is far better than all others. One can see that the skewness changes sign as $\Delta t$ crosses the optimal value $\Delta t^*$.
Using a similar argument to the one given in the previous section, one finds here that for $\Delta t \ne \Delta t^*$, the solution is only expected to be accurate (in the sense that the relative error is much smaller than $1$) within a spatial regime around $x=Ft$ of typical size
\be
\ell(t) \sim t^{2/3}
\ee
whereas for $\Delta t = \Delta t^*$, the relevant lengthscale is
\be
\ell(t) \sim t^{3/4} \, .
\ee

However, note that if the spatial discretization is very fine, $\Delta x\ll D/F$, then the optimal time step \eqref{DeltatSolRWD} becomes $\Delta t^{*}\simeq\Delta x^{2}/\left(6D\right)$, coinciding with that of the advection-free case.
Thus, at $\Delta x\ll D/F$, choosing $\Delta t = \Delta t^{*}$ is optimal, because for this choice the third and the fourth cumulants of $P(X^j)$ both (nearly) vanish, so the lowest nonvanishing cumulant will be the fifth. This leads to a further enhancement of the accuracy, encompassing a spatial regime whose typical size grows as
\be
\label{ellApproxAdvection}
\ell(t) \sim t^{4/5} \, .
\ee
 This appears to be the case in Fig.~\ref{fig:RWDs}  for the moderately small value $\Delta x=D/\left(3F\right)$.
 
 This analysis also suggests that, regardless of the force $F$, the choice $\Delta t = \Delta x^{2}/\left(6D\right)$ should yield good performance in terms of efficiency and accuracy, as long as $\Delta x\ll D/F$ (because the difference between this choice and the choice $\Delta t = \Delta t^*$ is very small). One may thus expect this choice $\Delta t = \Delta x^{2}/\left(6D\right)$ to perform quite well even if $F$ is not constant in space and/or time.

\section{Application: area swept by a Brownian particle in a potential $\sim|x|$}
\label{sec:Reflected}

As an application, we consider a model of Brownian motion in an external potential $V(x) = \mu|x|$, as described by the Langevin equation \cite{TSJ10, CBHJ13,DBK23}
\be
\label{LangevinOrig}
\dot{x}\left(t\right)=-\mu \, \text{sgn}\left(x\left(t\right)\right)+\sqrt{2D}\,\xi\left(t\right) \, .
\ee
where $\mu>0$ and $-\infty < x < \infty$. 
The particle starts at the origin, $x(t=0)=0$.
We study the full distribution $P\left(A;t\right)$ of the area
\be
\label{Adef}
A(t)=\int_{0}^{t}x\left(\tau\right)d\tau
\ee
swept by the particle up to time $t$.
This problem is studied theoretically in Ref.~\cite{Smith24Absx}, and our goal here is to study it numerically.
Using that $dA/dt = x$, the Fokker-Planck equation for the joint distribution $\mathcal{P}\left(x,A;t\right)$ of $x$ and $A$ at time $t$ is given by
\be
\label{FPAx}
\partial_{t}\mathcal{P}=-x\partial_{A}\mathcal{P}-\mu \, \partial_{x}\left[\text{sgn}\left(x\right)\mathcal{P}\right]+D\partial_{x}^{2}\mathcal{P} \, .
\ee
Eq.~\eqref{FPAx} is to be solved for $-\infty < x < \infty$ and $-\infty < A < \infty$ with the initial condition $\mathcal{P}\left(x,A;t=0\right)=\delta\left(x\right)\delta\left(A\right)$.
Once $\mathcal{P}\left(x,A;t\right)$ is found, one can marginalize over $x$ to obtain the area distribution:
\be
P\left(A;t\right)=\int_{-\infty}^{\infty}\mathcal{P}\left(x,A;t\right)dx \, .
\ee

In Ref.~\cite{Smith24Absx}, a variety of theoretical tools from large-deviation theory are applied to obtain the full distribution $P\left(A;t\right)$ (including its tails) approximately, in the long-time limit $t \to \infty$.
In the present work, we obtain $P\left(A;t\right)$ numerically by solving Eq.~\eqref{FPAx} using the finite-differences method. The numerical solution is computationally heavy, especially due to the fact that we are interested in obtaining $\mathcal{P}$ at long times and large $A$ (and, as a result, rather large $x$). As a result, one has no choice but to use relatively large values of the discretization parameters $\Delta x, \Delta A, \Delta t$, and it becomes of great importance to retain the best level of accuracy possible.

Let us now describe our numerical method. We rely on the results of Section \ref{sec:AdvectionDiffusion1D}, while noticing that in each of the regimes $x>0$ and $x<0$, Eq.~\eqref{LangevinOrig} describes a diffusing particle with a constant drift term ($F=-\mu$ and $F=\mu$ respectively). 
Let us describe our discretization method in terms of its corresponding stochastic dynamics that are discrete in $x$, $t$ and $A$. 
As above, we consider a random walker $\mathcal{X}(t)$ that, at every timestep $\Delta t$, jumps a distance $\Delta x$ to the right (left) with probability $p$ $(q)$, or, with probability $1-p-q$, stays in the same position. The optimal choices of $p,q$ and $\Delta t$ are given by Eqs.~\eqref{pBetter}, \eqref{qBetter} and \eqref{DeltatSolRWD}, respectively, where one must use $F=\mu$ ($F=-\mu$) if the position of the particle is to the right (left) of the origin%
\footnote{If the particle is exactly at the origin, we use $F=-\mu$ to calculate $p$ and $F=\mu$ to calculate $q$.}.
In addition, we consider a discrete-time process $\mathcal{A}(t)$ that, at every timestep $\Delta t$, jumps a distance of $\mathcal{X}(t)$. Note that $\mathcal{A}(t)$ takes discrete values, that are all given by integer multples of $\Delta A \equiv \Delta X \Delta t$.
The dynamics of the joint distribution of $\mathcal{X} = i\Delta X$ and $\mathcal{A}= j\Delta A$ at time $t=k\Delta t$ are described by a discrete version of Eq.~\eqref{FPAx}
\be
\label{RWxA}
\mathcal{P}_{i,j}^{k+1}=\left(1-p-q\right)\mathcal{P}_{i,j-i}^{k}+q\mathcal{P}_{i+1,j-\left(i+1\right)}^{k}+p\mathcal{P}_{i-1,j-\left(i-1\right)}^{k}\,,
\ee
where $i,j\in \mathbb{Z}$, $k = 0,1,2,\dots$, and $p$ and $q$ depend on the sign of $i$ as described above. The three terms on the right-hand side of Eq.~\eqref{RWxA} correspond to the three possible positions from which the random walker may reach the point $\mathcal{X} = i\Delta X$.

We implemented this solution with $D = 1$ and $\mu = 0.693$. Choosing $\Delta x = 1$, Eqs.~\eqref{pBetter}, \eqref{qBetter} and \eqref{DeltatSolRWD} yield
$\Delta t \simeq 0.1645$,
$p \simeq 0.114$ and $q \simeq 0.228$
at $x>0$ (at $x<0$ the values of $p$ and $q$ are exchanged). In order to probe the large-deviation regime, we used fairly large cutoff values,
$\mathcal{X}_{\max} = 80 \Delta x$ and
$\mathcal{A}_{\max} = {10}^5 \Delta A$,
and we were able to run the solution up to time $t=5\times{10}^4 \Delta t \simeq 8225$ within reasonable computational time.
The resulting $P(A;t)$ is plotted in \cite{Smith24Absx}, showing good agreement with the theoretical predictions obtained there.

\begin{figure*}[h]
\includegraphics[width=0.6\linewidth,clip=]{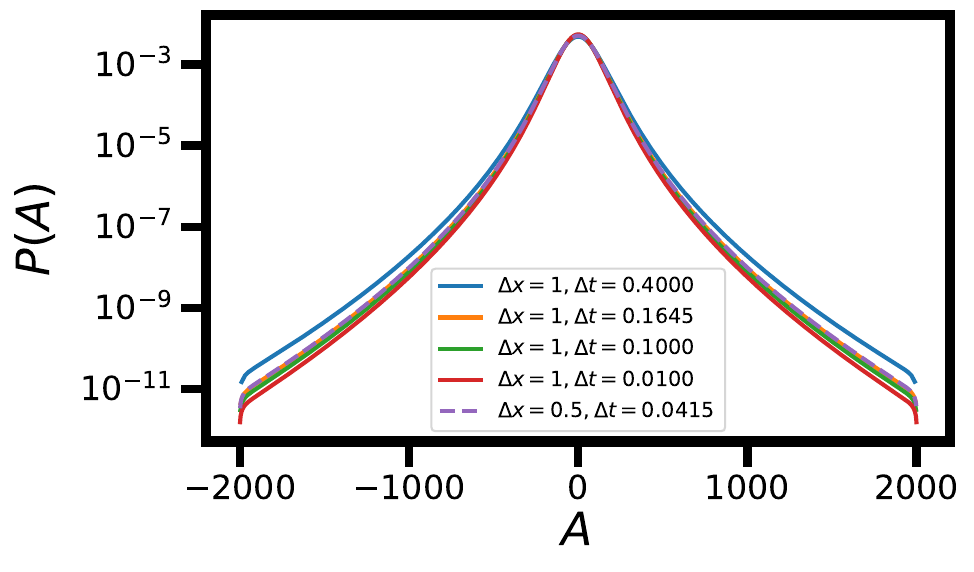}
\caption{Distributions $P(A;t)$ of the area \eqref{Adef} swept by a Brownian particle in an external potential $\mu|x|$, obtained from numerical solutions of the joint Fokker-Planck equation \eqref{FPAx} for the particle's position $x(t)$ and the area $A(t)$. Parameters are $D = 1$ and $\mu = 0.693$. The solid lines correspond to a spatial discretization of $\Delta x = 1$ and different choices of the temporal discretization, $\Delta t = 0.4, \Delta t^*, 0.1, 0.01$ where $ \Delta t^* = 0.1645$. We find that the choice $\Delta t = \Delta t^*$ yields the best accuracy, by comparing it with the dashed line, which corresponds to a numerical solution with a finer spatial discretization $\Delta x = 0.5$ and the corresponding optimal temporal discretization $\Delta t = 0.0415$.}
\label{fig:PAs}
\end{figure*}

 To test the dependence of the numerical accuracy on the chosen parameters, we also performed shorter runs, up to time $t=200$, with different choices of the time discretization, $\Delta t = 0.4, 0.1645, 0.1, 0.01$. We compared them with a more accurate numerical solution which we obtained by using a finer spatial discretization, $\Delta x = 0.5$, and the corresponding optimal time discretization $\Delta t = 0.0415$ from \eqref{DeltatSolRWD}. In all cases (i.e., for each choice of $\Delta x$ and $\Delta t$), we used the values of $p$ and $q$ from  Eqs.~\eqref{pBetter}, \eqref{qBetter}.
The resulting $P(A;t)$ are plotted in Fig.~\ref{fig:PAs}. Using the solution with $\Delta x = 0.5, \Delta t = 0.0415$ as a reference, we find that of all the solutions with $\Delta x = 1$, the choice $\Delta t = 0.1645$ yields the best accuracy, whereas the other choices yield an error that, in the large-deviation regime, is of order a factor of a few (this includes the choices  $\Delta t < \Delta t^*$ which are computationally more expensive).

\section{Higher dimension}
\label{sec:highDim}

In $d>1$, we can follow a similar approach as we did in $d=1$. Namely, we interpret the dynamics obtained from the finite-difference approximation as the evolution of the position distribution of a random walker in discrete space and time, and choose the parameters of the numerical solution such that the lowest cumulants coincide with those of Brownian motion (for simplicity, we consider a delta function initial position distribution and an infinite system).

Let us begin by considering the diffusion equation $\partial_{t}u=D\nabla^{2}u$ in $d=2$. A standard way to apply the finite-differences method is to approximate the Laplace operator by
\be
\label{LaplaceDiscrete}
\nabla^{2}u\left(x,y\right)\simeq\frac{u\left(x+\Delta\ell,y\right)+u\left(x-\Delta\ell,y\right)+u\left(x,y+\Delta\ell\right)+u\left(x,y-\Delta\ell\right)-4u\left(x,y\right)}{\Delta\ell^{2}} \, ,
\ee
where $\Delta\ell$ is the resolution of the spatial grid.
Using this in the diffusion equation, and approximating the time derivative by 
$\partial_{t}u\simeq\left(u|_{t+\Delta t}-u|_{t}\right)/\Delta t$,
 one obtains, in analogy to Eq.~\eqref{RW1D}, an equation that can be interpreted as describing the dynamics of a discrete random walker, that jumps with probability 
 $p=D\Delta t/\Delta x^{2}$
  a distance of $\Delta \ell$ to each one of the four directions $\pm \hat{x}, \pm \hat{y}$, or does not move with probability $1-4p$.
One may now try to choose $p$ such that the numerical error due to discretization is minimized.
As in the case $d=1$, one can choose $p=1/6$ so that the fourth cumulants of the $x-$ and $y-$ coordinates of the random walker vanish. However, it is not difficult to see that the $x$ and $y$ coordinates of such a particle are statistically dependent. Thus, the mixed correlation functions such as
$\left\langle x^{2}y^{2}\right\rangle -\left\langle x^{2}\right\rangle \left\langle y^{2}\right\rangle $
will not vanish, leading to nonvanishing mixed cumulants.

This problem is easily remedied if one uses instead the so-called locally one-dimension (LOD) method, in which each time step consists of two substeps, one in the $x$ direction and then one in the $y$ direction (these substeps correspond to the two parts $\partial_{x}^{2}u$ and $\partial_{y}^{2}u$ that together form the Laplacian).
The interpretation of these dynamics is as those of a random walker that, at each time step of duration $\Delta t$, first jumps to the left or right each with probability $p$ (or does not move, with probability $1-2p$) and then jumps up or down each with probability $p$ (or does not move, with probability $1-2p$).
Clearly, the $x$ and $y$ coordinates of such a random walker are exactly statistically independent.
One then follows the same analysis as for the $d=1$ case and chooses $p=1/6$, so that the fourth cumulants vanish.

\begin{figure*}[ht]
\includegraphics[width=0.48\linewidth,clip=]{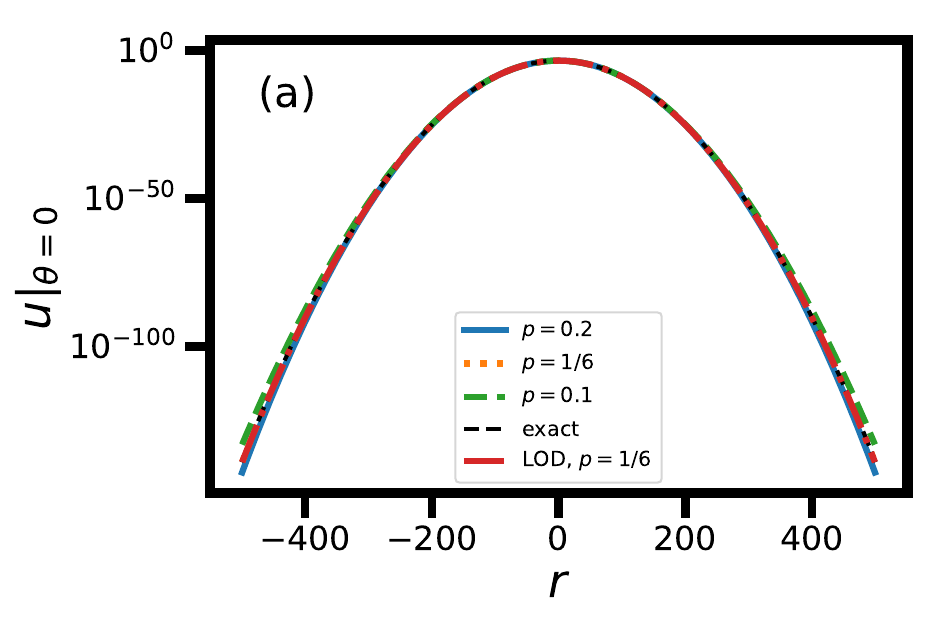}
\hspace{1mm}
\includegraphics[width=0.45\linewidth,clip=]{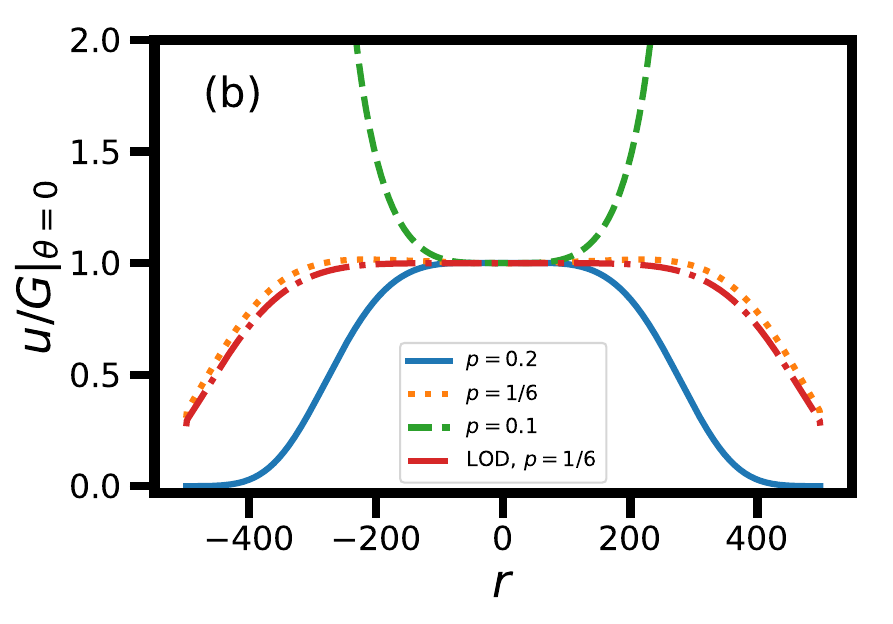}
\includegraphics[width=0.48\linewidth,clip=]{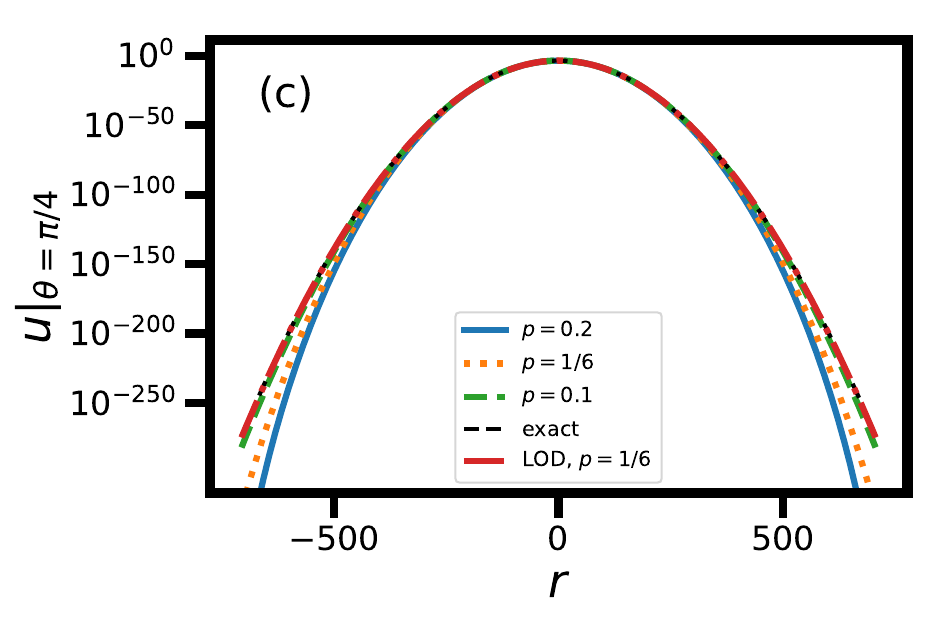}
\hspace{1mm}
\includegraphics[width=0.45\linewidth,clip=]{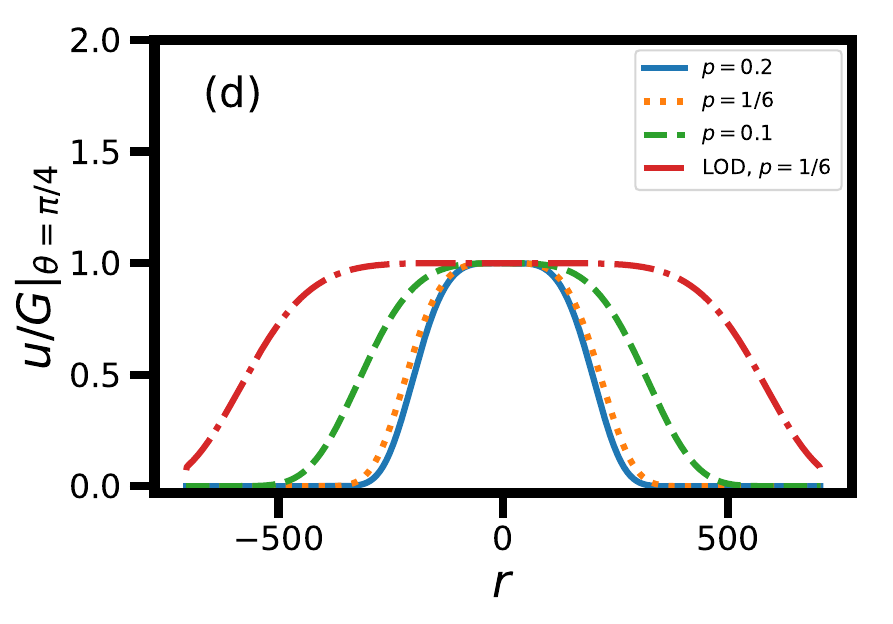}
\caption{(a) Numerical solutions $u(x,y,t)$ of the diffusion equation in $d=2$, obtained using the finite-differences method with the standard discretization of the Laplace operator \eqref{LaplaceDiscrete} and different values of $p=D\Delta t/\Delta x^{2}$, compared with the solution using the LOD method with $p=1/6$. Parameters are $D=\Delta x = 1$ and $t=200$. The solution is plotted as a function of the radial coordinate $r$ on the line $\theta=0$ where $\theta$ is the polar coordinate (i.e., the line $y=0$). Also plotted is the exact solution \eqref{Green2d}.
(b) The ratio between the numerical solutions and the exact solution \eqref{Green2d}. In this particular case ($\theta=0$), the choices $p=1/6$ with the standard discretization \eqref{LaplaceDiscrete} or with the LOD method both give excellent accuracy.
(c) and (d): The same as (a) and (b) but for the line $\theta = \pi/4$ (i.e., the line $x=y$). In (d) it is clearly seen that the LOD method is far more accurate than when using the standard discretization \eqref{LaplaceDiscrete}, even for $p=1/6$.}
\label{fig:RW2d}
\end{figure*}

In order to test this method, we apply it to the diffusion equation in $d=2$ equation with a delta-function initial condition, whose analytic solution is
\be
\label{Green2d}
u\left(x,y,t\right)=G\left(x,y,t\right)=\frac{1}{4\pi Dt}e^{-r^{2}/\left(4Dt\right)}\,,
\ee
where $r^2 = x^2 + y^2$. Note in particular that the $x$- and $y$-coordinates are statistically independent, because $u(x,y,t)$ may be decomposed as the product of the one-dimensional Green's functions \eqref{Green} acting on each coordinate separately.
As shown in Fig.~\ref{fig:RW2d}, the LOD method with $p=1/6$ is indeed far more accurate than all alternatives.

It is straightforward to show that the LOD method with $p=1/6$ leads to optimal accuracy for the diffusion equation in arbitrary dimension $d>1$, and that the dynamical lengthscale $\ell(t)$ over which the solution is expected to be accurate behaves as in $d=1$, see Eq.~\eqref{ellGood}.
Similarly, for the advection-diffusion equation in $d>1$, the LOD method with discretization parameters chosen as described in section \ref{sec:AdvectionDiffusion1D} is expected to lead to optimal accuracy.

\section{Summary and discussion}
\label{sec:conclusion}

To summarize, we studied the optimal choice of parameters for numerical solutions of the diffusion equation in arbitrary dimension $d$ using the finite-differences method. By interpreting the numerical solution as the dynamics of the distribution of the position of a random walk in discrete time and space, we found that
the choice of the time discretization that gives optimal accuracy is $\Delta t = {\Delta \ell}^2 / (6D)$, where $\Delta \ell$ is the spatial discretization,
thus recovering the known result of truncation error analysis.
Rather counterintuitively, if $\Delta t$ is decreased beneath this value, the accuracy worsens. The accuracy in large-deviations regimes, in which the concentration of particles is very small, is especially sensitive to the choice of $\Delta t$. In $d>1$ we found that combining the choice of $\Delta t$ reported above with the LOD method provides much higher accuracy than that which is obtained from a standard discretization of the Laplace operator.
We extended our results to the case of nonzero advection, and found in this case the parameters $p,q,\Delta t$ for which optimal accuracy is obtained. 
 These parameters too coincide with the result obtained from truncation error analysis.
We then illustrated the method by applying it to compute the distribution of the position of the area swept by a Brownian particle trapped by an external potential $\sim |x|$.


We used the quantitative tools from large-deviation theory to analyze the accuracy of the numerical solution to the diffusion and advection-diffusion equations with the finite-differences method, and fine-tune the parameters of this method to optimize its accuracy.
It would be interesting to see if this approach can be extended to optimize other numerical schemes (for instance, implicit methods \cite{GG78}) and/or to the numerical solution of related equations  such as reaction-advection-diffusion equations \cite{ALV96}. 
In particular, one could consider the case  in which the forcing $\vect{F}$ and/or diffusion $D$ are time- or space-dependent, or even depend on the concentration $u$ itself. It appears unlikely that it would be possible to obtain a method that performs as well as in the relatively simple case of constant $\vect{F}$ and $D$ considered here. However, in the particular case where $D$ is constant, and the spatial discretization $\Delta x$ is chosen to be sufficiently small, one might still expect that choosing $\Delta t = \Delta x^{2}/\left(6D\right)$ would result in good performance  in terms of efficiency and accuracy, as explained shortly after Eq.~\eqref{ellApproxAdvection}.
This analysis could prove useful for studying, e.g., large-deviation statistics in the Ornstein-Uhlenbeck process \cite{NT18, Smith22OU, NT22}, in which there are theoretical predictions which still await numerical verification.

Finally, numerical investigation of large-deviation statistics is a notoriously difficult task, due to the extremely low probabilities of the events involved \cite{Hartmann2002,	TL09, HG11, touchette2011, Claussen2015, Hartmann2018, GV19, Hartmann2019LIS, Hartmann2019,Hartmann2020,HartmannMS2021, CBG23, SmithChaos22, SmithFarago22, NT22}.
 We pointed out that an optimal choice of parameters for the finite-difference discretization of the advection-diffusion equation can have a huge effect (many orders of magnitude) in the regions of space and time for which the concentration $u$ is small. These regions are often those that are of crucial importance when numerically analyzing large-deviation statistics.
It would be useful to apply our results to  numerically study  large-deviation properties of diffusing particles with or without drifts (either when solving diffusion equations numerically \cite{Smith24Absx}, or in Monte-Carlo simulations of Brownian motion \cite{Claussen2015}). The example given in the present work (area swept by a trapped Brownian particle) together with Ref.~\cite{Smith24Absx} is a first step in this direction.

\subsection*{Acknowledgments}

I thank Lucianno Defaveri and Eli Barkai for helpful discussions. I acknowledge support from the Israel Science Foundation (ISF) through Grant No. 2651/23.

\appendix

\medskip

\section{Truncation error analysis}
\label{app:truncation}

Consider the one-dimensional diffusion equation $\partial_{t}u=\partial_{x}^{2}u$, where we chose $D=1$ for convenience.
By Taylor expanding $u(x,t)$, one can calculate the error due to the finite-difference approximations of the time and space derivatives,
\bea
\label{utTaylor}
\frac{u\left(x,t+\Delta t\right)-u\left(x,t\right)}{\Delta t}&=&\partial_{t}u\left(x,t\right)+\frac{\Delta t}{2}\partial_{t}^{2}u\left(x,t\right)+\dots,\\
\label{uxTaylor}
\frac{u\left(x+\Delta x,t\right)-2u\left(x,t\right)+u\left(x-\Delta x,t\right)}{\Delta x^{2}}&=&\partial_{x}^{2}u\left(x,t\right)+\frac{\Delta x^{2}}{12}\partial_{x}^{4}u\left(x,t\right)+\dots.
\eea
Now, since $u$ satisfies the diffusion equation, $\partial_{t}^{2}u = \partial_{x}^{4}u$. One therefore finds that if $p=\Delta t/\Delta x^{2} = 1/6$, the two error terms on the right-hand sides of Eqs.~\eqref{utTaylor} and \eqref{uxTaylor} are equal, and thus this choice of $p$ gives maximal accuracy (the numerical error will be only due to higher-order terms in the Taylor series), as stated in the main text.

\end{document}